\documentstyle[aps,prl,multicol,psfig]{revtex}
\renewcommand{\narrowtext}{\begin{multicols}{2} \global\columnwidth20.5pc}
\renewcommand{\widetext}{\end{multicols} \global\columnwidth42.5pc}

\begin{document}
\def\CC{{\rm\kern.24em \vrule width.04em height1.46ex depth-.07ex
\kern-.30em C}}
\def\P{{\rm I\kern-.25em P}}
\def\RR{{\rm
         \vrule width.04em height1.58ex depth-.0ex
         \kern-.04em R}}

\draft
\title{  On the Entangling Power of  Quantum Evolutions}
\author{Paolo Zanardi$^{1,3}$, Christof Zalka$^1$ and Lara Faoro$^{2,3}$ }

\address{
 $^1$ Institute for Scientific Interchange  (ISI) Foundation, 
Viale Settimio Severo 65, I-10133 Torino, Italy\\
$^2$ Dipartimento di Fisica Politecnico di Torino, Viale Duca degli Abruzzi 24, 
I-10129 Torino, Italy\\
$^3$ Istituto Nazionale per la Fisica della Materia (INFM) }
\date{\today}
\maketitle

\begin{abstract}
We analyze the entangling capabilities of  
unitary transformations $U$ acting on a bipartite $d_1\times d_2$-dimensional quantum system.
To this aim we introduce an entangling power  measure  $e(U)$ 
given by the mean linear entropy produced 
acting with $U$ on a given distribution of pure product states.
This measure admits a natural interpretation in terms of quantum operations.
For a uniform distribution explicit analytical results are obtained
using group-theoretic arguments.
The  behavior  of the features of $e(U)$ as the subsystem dimensions $d_1$ and $d_2$
are varied is studied both analytically and numerically.
The two-qubit case $d_1=d_2=2$ is argued to be peculiar.
 
\end{abstract}
\pacs{PACS numbers: 03.67.Lx, 03.65.Fd}
\narrowtext
From the  beginning it has been argued  that  entanglement is one of the crucial ingredients
that allows  Quantum  Information processing \cite{QC} to outperform, for certain tasks,
any classically operating device.  
In this sense entanglement represents  a uniquely quantum resource
whose production is  a sort of elementary prerequisite for any Quantum Computation (QC). 
Such a basic task is accomplished by unitary transformations $U$ i.e., quantum evolutions
acting on the state-space of the multi-partite system that describe non-trivial interactions
between the degrees of freedom of the different subsystems.
Even though almost all the unitaries satisfy this latter requirement \cite{UG},
it is quite natural to ask how different $U$'s are efficient, according to some criterion  to be specified,
as entanglers, and then by using such a criterion to analyze the full manifold of bi-partite quantum evolutions.
  
In this paper we address this issue by introducing over the space of bi-partite unitaries
a measure for their {\em entangling power}.
This is done by considering how much entanglement is produced by $U$
{\em on the average} acting on a given distribution of 
unentangled  quantum states.
The kind of situation we have in mind is a procedure for entanglement production
in which one randomly generates product states (the "cheap" resource )
according to some  probability distribution $p$ and then applies the transformation $U.$
The average entanglement obtained with the above scheme will be 
our  measure  $e_p(U)$ of the quantum evolution $U.$

It  is important to stress  that  these $U$'s 
can  represent different objects,
both from the logical and physical point of view.
Some prototypical instances are given by:
a) A quantum computation using a pair of quantum registers.
Here the  entangling power measure  will quantify
how the computation $U$ is efficient in making
the first  (say memory) and the second (say computational)
registers entangled. This kind of entanglement, that represents mutual information
between the two registers, has been recently
proved to play a role in QC viewed as a communication process \cite{bose}.
b) The global evolution  of a system plus its environment.
 In this case $e_p(U)$ measures the {\em decohering}  
power of the system-environment coupled evolution $U.$
Engineering weak decoherence then amounts to design an optimal
$U$ with respect to the criterion of {\em minimal} entangling power.
c) A single two-subsystem e.g., two-qubits, gate in a quantum-network.
Now the entangling $U$ are the two-qubit gates needed
to get  universal QC \cite{UG}.

To   formalize our  setting let us consider a bipartite quantum system with state space 
${\cal H}={\cal H}_1\otimes {\cal H}_2$
where dim$\,{\cal H}_i=d_i\,(i=1,2)$
and $U\in {\cal U}({\cal H})\cong U(d_1\,d_2).$
If $E$ is an entanglement measure over ${\cal H}$ we define the {\em entangling power}
of $U$ (with respect to $E$)
as
\begin{equation}
e_p( U) :=  \overline{ E( U\,|\psi_1\rangle\otimes|\psi_2\rangle)}^{\psi_1, \psi_2}
\label{power}
\end{equation}
where the bar  denotes the average 
over to all the {\em product} states $|\psi_1\rangle\otimes|\psi_2\rangle.$
distributed according some probability density $p(\psi_1, \psi_2)$
over the manifold of product states.

We shall use as entanglement measure of $|\Psi\rangle\in{\cal H}$
the {\em linear entropy}
\begin{equation}
E(|\Psi\rangle) := 1- \mbox{tr}_1 \rho^2,\;
\rho:= \mbox{tr}_2 |\Psi\rangle\langle\Psi|.
\label{lin-en}
\end{equation}
This quantity measures the  purity of the reduced density matrix $\rho,$ it
can be regarded as a kind of "linearized" version of the von Neumann entropy $S(\rho) =-\mbox{tr}\,\rho\ln\rho,$
which  is known to provide the essentially unique measure of entanglement for bi-partite pure quantum states.
One has that $0\le E(|\Psi\rangle)\le 1-1/d$ where the lower (upper) bound is reached
iff $|\psi\rangle$ is a product state (maximally entangled).
The  measure (\ref{lin-en}) has, with respect to $S(\rho),$ the definite advantage
of being a {\em polynomial} in $|\psi\rangle.$

Now we introduce some notations.
We shall denote by $T_{ij},\, (i,j=1,\ldots,4)$ the transposition
between the $i$-the and the $j$-th factor of
${\cal H}^{\otimes\,2}:= {\cal H}_2\cong (\CC^{d_1}\otimes\CC^{d_2})\otimes(\CC^{d_1}\otimes\CC^{d_2}).$
Notice that $T_{12}$ and $T_{34}$ are well defined elements of ${\cal U}({\cal H}^{\otimes\,2})$
only when $d_1=d_2,$ in this latter case such operators will be referred to as {\em swaps}.
Moreover -- when ${\cal H}_i\cong {\cal H}_j$ -- 
one defines the projectors $P^{\pm}_{ij}:= 2^{-1}(\openone \pm T_{ij})$
over the totally symmetric (antisymmetric)  subspaces of ${\cal H}_i\otimes{\cal H}_j,$
the latter being thought of as embedded in ${\cal H}^{\otimes\,2}.$
 The space End $ ({\cal H}^{\otimes\,2})$ is endowed with  the Hilbert-Schmidt scalar product
$<A,\,B>:=\mbox{tr} (A^\dagger\,B).$ Finally with ${\cal S}({\cal H})$ we shall denote
the space of density matrices over ${\cal H}.$

{\bf{Proposition 0}}
The entangling power (\ref{power}) is given by
\begin{eqnarray}
e_p(U)=2\,\mbox{tr}\,[ U^{\otimes\,2}\,\Omega_p\,  U^{\dagger\otimes\,2}\,P_{13}^-],
\label{ep-general}
\end{eqnarray}
where  $\Omega_p :=\int d\mu(\psi_1,\psi_2) (|\psi_1\rangle\langle\psi_1|\otimes
 |\psi_2\rangle\langle\psi_2|)^{\otimes\,2}
\in{\cal S}({\cal H}^{\otimes\,2})$ and $d\mu$ denotes
the measure over the product state manifold induced by the probability distribution $p(\psi_1, \psi_2).$

{\em Proof.}
 Let  us observe that Eq. (\ref{lin-en}) can be written in a linear form
using the identity tr$\,[ (A \otimes B)\,T]= \mbox{tr}\, (A\,B)$ where $T$ is the swap.
Then $E(|\Psi\rangle)=1-\mbox{tr}\,( |\Psi\rangle\langle\Psi|^{\otimes\,2}\,T_{13}).$
Form this remark and the definition (\ref{power}) it follows immediately (\ref{ep-general})$\hfill\Box$

The result above  express $e_p(U)$ as the expectation value over $\Omega_p$ of the
 positive operator $2\,U^{\dagger\otimes\,2}\,P_{13}^-\,U^{\otimes\,2}.$
This latter operator can be viewed as the {\em effect} associated to the completely positive (CP) \cite{KRA}  map $\Phi_U$ on 
${\cal S}({\cal H}^{\otimes\,2})$ 
given by 
$
\Phi_U\colon\Omega\mapsto 2\,P_{13}^-\,U^{\otimes\,2}\,\Omega\,U^{\dagger\otimes\,2}\,P_{13}^-. 
$
This remark allows us to interpret the entangling power (\ref{ep}) as probability of success
of a two-party (A and B) quantum protocol  (see Fig. 1).
Suppose $A$ ($B$) owns spaces ${\cal H}_1$ and ${\cal H}_3$ (${\cal H}_2$ and ${\cal H}_4$)

 a) A and B generate pairs of states $|\psi_1\rangle\otimes|\psi_2\rangle$
  according the distribution probability $p(\psi_1, \psi_2)$ [$\Omega_p$ is prepared]  
b) Apply to each member of the pair the  joint transformation $U$ [action of $U^{\otimes\,2}$]
c) Perform a projective measurement of $\sqrt{2}\,P^-_{13}.$
\begin{figure}
\unitlength1mm
\ \par\noindent
\begin{picture}(80,55)
\put(0,0){\psfig{figure=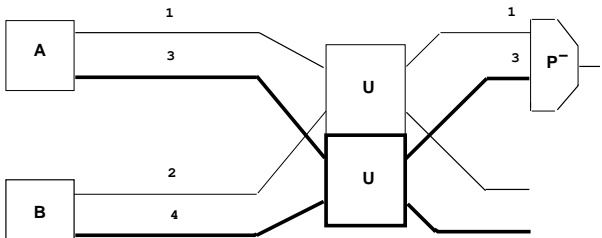,width=80mm}}
\end{picture}
\caption{\label{fig1}
 Scheme of the two-party protocol realizing the operation $\Phi_U.$
}
\end{figure}
Eq. (\ref{ep-general}) nicely displays several properties
required for {\em any} entangling measure for bi-partite unitary evolutions.
i) $e_p(U_1\otimes U_2\,U)=e_p(U)\, (U_i\in U(d_i))$.
   Indeed from the $U(d_1)$-invariance of $P_{13}^-$ one finds 
 $e_p(U_1\otimes U_2\,U)= 2\,\mbox{tr}\,[U^{\otimes\,2}\,\Omega_p\,U^{\dagger\otimes\,2}
\,(U_1\otimes U_2)^{\dagger\otimes\,2}\, P_{13}^-\,(U_1\otimes U_2)^{\otimes\,2}
]=2\,\mbox{tr}\,[U^{\otimes\,2}\,\Omega_p\,
U^{\dagger\otimes\,2}\,(U_1)_{13}^\dagger\,P_{13}^-\,(U_1)_{13}]=e_p(U).$
Where $(U_1)_{13}:= U_1\otimes\openone\otimes U_1\otimes \openone.$
ii)  When $d_1=d_2,$ by denoting with  $T$  the transposition
between the two factors of $\cal H,$
one has  $e_p(T\,U)=e_p(U).$ This stems from $T^{\otimes\,2}\,P_{13}^-\,T^{\otimes\,2}=P_{24}^-.$
This leaves $e_p(U)$ unchanged, indeed
 this label change amounts simply to the replacement tr$_1\leftrightarrow$ tr$_2$  in Eq. (\ref{lin-en}).
Since, for pure states, the two reduced density matrices are isospectral the linear entropy is unchanged.
Moreover if $\Omega_p$ is swap invariant i.e., $p(\psi_1, \psi_2)= p(\psi_2, \psi_1)$
one also has $e_p(U\,T)=e_p(U).$  
iii) One has $e_p(\openone)=0.$ This simply because $\Omega_p\,P_{13}^-=0.$
Indeed, form the definition (\ref{ep-general}) one has  $\Omega_p\,T_{13}= \Omega_p.$
iv) From the previous remarks it follows that the entangling power
     is constant along the orbits in ${\cal U}( {\cal H})$
     of the {\em left} action  of the subgroup
     of the  bi-local operations $U_1\otimes U_2.$
      In particular $e$ vanishes on all the elements of such a group.
     In the symmetric case $d_1=d_2$ the group is extended by the swap $T.$

Different distributions $p(\psi_1,\,\psi_2)$
would result in very  different $e(U).$
An extreme example of this obvious remark is provided
by transformations  $U$ that simply permutes elements of a given basis $|i\rangle\otimes|j\rangle$
of ${\cal H}$. If $p$ is supported just on this basis the associated $e_p(U)$ vanishes identically,
while we shall show later for a different probability distribution that  $U$'s can  even be  maximally entangling.
Another example is given in the context of the case b) mentioned in the introduction.
Suppose $U$ admits a decoherence-free subspace ${\cal C}\subset {\cal H}_1$ \cite{EAC}
the if $p$ is, for any $|\psi_2\rangle,$  supported in $\cal C$ then again $e_p(U)=0.$

From now on  we  focus  on the case in which  $p$ is the {\em uniform} distribution $p_0$.
With this term we refer to the unique 
 $U(d_1)\times U(d_2)$-invariant probability distribution i.e.,
$p(\psi_1, \psi_2)=p(U_1\,\psi_1,\, U_2\,\psi_2).$
When  all the product state are considered to be equally easy to  
be  prepared, this latter assumption on $p$ is quite natural from
the physical point of view \cite{worth}.
Moreover, in view of its symmetry, the uniform $p$
will result in a great computational simplification  
that will allow  for an explicit analytical evaluation of the average over the product-state
manifold that appears in Eq. (\ref{power}).

Let us begin by proving an easy group-theoretic Lemma
that will play an essential technical role in the following.

{\bf {Lemma}}
$\Omega_{p_0} =4\,C_{d_1}\,C_{d_2} \, P^+_{13}\,P^+_{24},\;
C_d^{-1} := d\,(d+1).$ 

{\em Proof.}
Since the uniform distribution factorizes 
we can consider separately the average $\omega_{13}$ with respect to $|\psi_1\rangle$
( on the first and the third factor of ${\cal H}^{\otimes\,2}$)
and the one  $\omega_{24}$ with respect to $|\psi_2\rangle$
( on the second  and the fourth  factor of ${\cal H}^{\otimes\,2}$)
then $\Omega_{p_0}=\omega_{13}\,\omega_{24}.$
Let first observe that in view of definition (\ref{ep-general}) one has 
that  $\Omega_{p_0}$ is supported in  $P_{13}^+\, P_{24}^+\,{\cal H}^{\otimes\,2}$
i.e., $\Omega_{p}$ is symmetric under the exchange of the first (second)
and the third (fourth) factor.
Moreover since the uniform distribution is $U(d_1)\times U(d_2)$ 
invariant one has $[U_{1}^{\otimes\,2},\, \omega_{13}]=0, \,\forall U_1\in U(d_1),$ 
and analogously for $\omega_{24}.$
Since the $U_1^{\otimes\,2}$'s act on the totally symmetric subspace irreducibly,
it follows from the above commutation relation and the Schur Lemma \cite{CORN}
that $\omega_{13} = 2\,C\, P_{13}^+.$
The normalization constant is found  by the condition tr$\, \omega_{13}=1.$
Reasoning in the same way for $\omega_{24}$ one gets 
the desired result.
$\hfill\Box$

{\bf {Proposition 1}}
The entangling power (\ref{power}), with respect the uniform distribution, is given by

\begin{eqnarray}
e_{p_0}(U) &=& 1- C_{d_1}\,C_{d_2}\sum_{\alpha=0,1} I_\alpha(U)\nonumber \\
I_\alpha(U) &=& t(\alpha)
 + <U^{\otimes\,2}\, (T_{1+\alpha, 3+\alpha})\,U^{\dagger\otimes\,2},\,T_{13}>,
\label{ep}
\end{eqnarray}
where $t(\alpha):=\mbox{tr}\, T_{1+\alpha, 3+\alpha}.$

{\em Proof.}
It is just a calculation.
Insert $\Omega_{p_0}= C_{d_1}\,C_{d_2}\, (\openone +T_{13})\,(\openone +T_{24})$
in Eq. (\ref{ep-general}). Notice that one has
tr $ T_{13} =d_1\,d_2^2,\, \mbox{tr}\,T_{24}=d_1^2\,d_2.$
$\hfill\Box$

From the relations  $[ T_{1+\alpha, 3+\alpha},\, (U_1\otimes U_2)^{\otimes\,2}]=0,\,(\alpha=0,1)$ 
it follows that {\em both} the functions $I_0$ and $I_1$
are invariant under the two-sided i.e.,left and right, action of  
bi-local unitaries e.g., $I_1(U)= I_1(U\,U_1\otimes U_2).$ 
Moreover  in the symmetric case $d_1=d_2$ 
 it follows  from the above that  the entangling power (\ref{ep}) can be written in a manifestly swap 
invariant form
$ e_{p_0}(U)=1-C_d^2 \sum_{i=0}^1 I_d(T^i\,U),$ where 
$I_d(U) := d^3 + <U^{\otimes\,2},\, T_{13}\, U^{\otimes\,2}\, T_{13} >.$
\begin{figure}
\unitlength1mm
\ \par\noindent
\begin{picture}(80,55)
\put(0,0){\psfig{figure=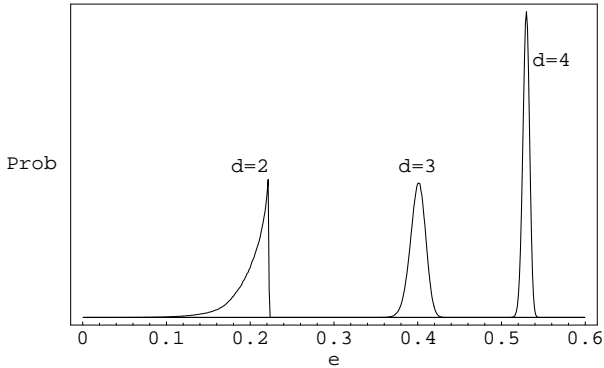,width=80mm}}
\end{picture}
\caption{\label{fig2}
Distribution of the probability density $q(e)$ for $d=2, 3 4.$
}
\end{figure}
The entangling power $e$ defines a random variable over ${\cal U}({\cal H})$ 
if the latter, endowed with the Haar measure, is considered  as a   probability space.
Therefore it 
makes sense  to consider the associated density of probability  distribution $q(e).$
Moreover, since the the manifold of unitary transformations over $\cal H$   is compact,
the obviously continuous mapping  $U\mapsto e(U)$
must achieves  extrema, in particular $\exists \bar U\in {\cal U}({\cal H})\colon e(\bar U) =\max_U e(U).$
Such maximally entangling $\bar U$'s will be referred to as {\em optimal}.
In Fig. 2 are reported the $q(e)$'s obtained numerically  for the cases $d\times d$ 
with $d=2,3,4.$
While in the cases $d\ge 2$  the function $q(e)$ vanishes on  both  the lower and the upper sides
of the allowed range of $e,$
it it remarkable that the two-qubit case $d=2$ shows a peculiar feature:
$q(e)$ is  a {\em monotonic}  function of $e.$
This implies that most of the two-qubit gates  $U$ correspond to nearly optimal ones.
Moreover, as will be discussed later in details, the entangling power of optimal $U$'s does not correspond
to an upper bound that is instead reached by  all the other cases for $d.$ 
In this sense the prototypical quantum information case of two qubits
is quite singular.
A first very natural question is how on the average an operator is entangling
i.e., the mean of the $q(e).$

{\bf Proposition 2} 
The average of the entangling power $e_{p_0}(U)$ over $U(d_1\,d_2)$ is given by
\begin{equation}
\overline{e_{p_0} (U)}^U = \frac{(d_1-1)\,(d_2-1)}{d_1\,d_2+1}
\label{mean}
\end{equation}
{\em Proof.} 
To prove Eq. (\ref{mean}) we first notice that, in view of definition (\ref{ep}),
to compute the mean of the entangling power amounts to compute the average
of the entanglement measure over {\em all} the states $|\Psi\rangle\in{\cal H}$
( not just over the product states).
The trace of the square of the reduced density matrix of $|\Psi\rangle$ is given by
$ \mbox{tr}\,(|\Psi\rangle\langle\Psi|^{\otimes\,2}\, T_{13} ) .$ 
Now we take the average with respect $|\Psi\rangle$ using again Lemma  
$ [d_1\,d_2 ( d_1\,d_2 +1)] ^{-1}\mbox{tr} [ (\openone + T_{13}\,T_{24} ) \, T_{13} ]= [d_1\,d_2^2+ d_1^2 ]\,
[d_1\,d_2 ( d_1\,d_2 +1)]^{-1}.$
This expression inserted in the definition of the entanglement measure proves Eq. (\ref{mean}).
$\hfill\Box$

For proving bounds on the entangling power (\ref{power}) 
it is  useful to 
consider one of the states, say $|\psi_2\rangle,$  of the input
product as fixed.
 In this case a pair of CP-maps associated with  $U$ are naturally defined.
Indeed one has
(explicit dependence on $U$ and $|\psi_2\rangle$ is omitted)
$\Phi\colon {\cal S}({\cal H}_1)\mapsto {\cal S}({\cal H}_1)\colon
\rho\mapsto\sum_{i=1}^{d_2} A_i\,\rho\,A_i^\dagger$ and
$\tilde  \Phi\colon {\cal S}({\cal H}_1)\mapsto {\cal S}({\cal H}_2)\colon
\rho\mapsto\sum_{i=1}^{d_1} \tilde A_i\,\rho\,\tilde A_i^\dagger
$
where $ A_i\colon {\cal H}_1\mapsto {\cal H}_1$ and $\tilde  A_j\colon {\cal H}_1\mapsto {\cal H}_2$ are given
by
$
A_i := \langle j|\,U\,|\psi_2\rangle ,\quad(j=1,\ldots,d_2) \nonumber \\
\tilde A_i := \sum_{j=1}^{d_2} |j\rangle\langle i|\,A_j,\quad (i=1,\ldots,d_1).
$
Therefore one can also define, for fixed $|\psi_2\rangle\in{\cal H}_2$
the (partial) entangling power of $U$ as
$
\tilde e_p(\Phi):=\overline{E(\Phi\,|\psi\rangle\langle\psi|)}^{\psi}.
$
Notice  that the equation above can also be  written 
is the form (\ref{ep-general}) with a special choice for $p(\psi_1, \psi_2)$ i.e., with
$\Omega_p = \int \mu(\psi_1) (|\psi_1\rangle\langle\psi_1|\otimes
 |\psi_2\rangle\langle\psi_2|)^{\otimes\,2}.$ 
The definition of  $\tilde e_p(\Phi)$ of course makes sense for general CP-maps, in this case
the expression for $\tilde e_p(\Phi)$ analogous to Eq. (\ref{ep-general}) is given by
$\tilde e_p(\Phi)= 2\,\mbox{tr}\,[\Phi^{\otimes\,2} (\omega_p)\,P_{13}^-]$
where  $\omega_p:=\int d\tilde \mu(\psi)\, |\psi\rangle\langle\psi|^{\otimes\,2}
\in{\cal S}({\cal H}_1^{\otimes\,2}).$

{\bf{Proposition 3}}
The entangling power of the CP-map $\Phi\colon {\cal S}({\cal H}_1)\mapsto {\cal S}({\cal H}_1)$
with respect to the uniform distribution is given by
\begin{equation}
\tilde e_{p_0}(\Phi)=1- C_{d_1}\,(\mbox{tr}_2\,\tilde X^2+ \mbox{tr}_1 X^2)
\label{ep0-cp}
\end{equation}
where $X:=\sum_{j=1}^{d_2} A_i\,A_i^\dagger$ and $\tilde X:=\sum_{j=1}^{d_1} \tilde A_i\,\tilde A_i^\dagger.$

{\em {Proof}.}
One has   that $\mbox{tr}_1\,\rho^2$
is given by $\sum_{i,j=1}^{d_2} \mbox{tr}_1\, (A_i\,|\psi_1\rangle\langle\psi_1|
 \,A_i^\dagger A_j\,|\psi_1\rangle\langle\psi_1| \,A_j^\dagger ),$
this last expression can be rewritten as 
$\sum_{i,j=1}^{d_2}
\mbox{tr}_1\, [ (A_j^\dagger\,A_i)\otimes (A_i^\dagger\,A_j)\,|\psi_1\rangle\langle\psi_1|^{\otimes\,2}
].
$
Now we perform the average with respect $|\psi_1\rangle;$ using Lemma 
$\omega_{p_0} = C_{d_1}\,(\openone + T).$
Using again the identity tr$\,[ (A \otimes B)\,T]= \mbox{tr}\, (A\,B)$ one gets
\begin{eqnarray}
C_{d_1}^{-1}\overline{\mbox{tr}_1\,\rho^2}^{\psi_1} = \sum_{i,j=1}^{d_2}
 |\mbox{tr}_1 (A_j^\dagger\,A_i)|^2 +\mbox{tr}_1\, (\sum_{1=}^{d_2} A_i\,A_i^\dagger)^2
\label{eq}
\end{eqnarray}
[Notice that the two terms in the equation above corresponds to the $I_\alpha$'s
in Eq. (\ref{ep}).]
It is then straightforward algebra to check that the first term in the equation above
can be written as $ \mbox{tr}_2\,\tilde X^2$ $\hfill\Box$

We  now provide bounds on the entangling power.
We assume that $d_1\le d_2.$

{\bf{Proposition 4} } For any  $U\in{\cal U}({\cal H})$ one has
\begin{equation}
 0 \le e_{p_0}(U) \le \frac{d_2-d_2/d_1}{d_2+1}.
\label{bound} 
\end{equation}

{\em Proof.} 
The lower bound is obvious in view of the definition (\ref{ep}), it is achieved
by all the unitaries obtained composing bi-local transformations of $U(d)\times U(d)$
with the swap.
Let us consider first the operator $X$ in Eq. (\ref{ep0-cp}).
By denoting with $\Phi$ the CP-map associated with  the $A_j$'s one has $\Phi(\openone/d_1)=X/d_1$
and then $1- \mbox{tr}_1 \,\Phi(\openone/d_1)^2 \le 1-1/d_1$ (general bound on linear entropy) 
it follows that  $d_1\le \mbox{tr}_1\,X^2.$  
This latter inequality provides a bound on the second  term of Eq. (\ref{ep0-cp}).
Reasoning in the same way with the operator $\tilde X$
and the associated CP-map $\tilde\Phi$ one finds for the first term of (\ref{ep0-cp})
the lower bound  $d_1^2/d_2.$
Putting these two results together, and in view of the assumption $d_1\le d_2,$ inverting 
$d_1$ with $d_2$ one gets the desired result (\ref{bound}) $\hfill\Box$

Another issue is to understand whether  the upper bound in Eq. (\ref{bound}) is achieved
by an optimal  unitary transformation $U.$
As we shall show in the following the answer seems to be affirmative for $d>2,$
Let us stress that this is not obvious at all in that
the upper bound (\ref{bound}) has been obtained by providing separate bounds
on the two  terms appearing in $e_{p_0}(U).$
Notice that the proof of Proposition 3 allows us to state the condition on $U$ in order to saturate the bound 
(\ref{bound}) as :
for any initial state $|\psi\rangle\in{\cal H}_1$ the associated CP-maps $\Phi$ and $\tilde \Phi$
(depending both on $|\psi\rangle$ and $U$) 
must be unital i.e., they map totally mixed states onto totally mixed states.
It  might well be that no $U$'s yields unitality for both CP-maps   at once.
This in fact turns out numerically to be the case for $d=2,$ in which  one has that the optimal
$U$'s are such that 
[see Fig. 2] $ e_{p_0}(U) = 2/9 < 1/3.$
An   optimal  operators for qubit is given (not surprisingly) 
by the controlled-not $U : =|0\rangle\langle 0|\otimes\openone+|1\rangle\langle1|\otimes X,$
where $X:= |0\rangle\langle 1|+|1\rangle\langle 0|.$
More interestingly   the operators
providing a natural $d$-dimensional generalization of the controlled-not
 are in general {\em not} optimal.
This is shown by the following calculation.

Let us   consider, for $d_1=d_2,$ 
$U=\sum_{\alpha=1}^d |\alpha\rangle\langle\alpha| \otimes U_\alpha,$
 where
the $|\alpha\rangle$'s are a $d$-dimensional orthonormal basis 
and the $U_\alpha$'s are unitaries  which, without any loss of generality can be taken to be 
orthogonal with respect to the Hilbert-Schmidt scalar product i.e.,
 $< U_\alpha,\,U_\beta>= d\,\delta_{\alpha, \beta}.$
By using Eq. (\ref{ep}) is easy to prove that for these unitaries one has
$e_{p_0}(U) =d\,(d-1)/(d+1)^2,$
that is $d/(d+1)$ times smaller than the bound $(\ref{bound})$ and, for $d>2$ even smaller.

We  also performed numerical 
maximization   of $e_{p_0}(U)$  trying to find exact expressions 
for the optimal unitary transformations. 
 For $d_1=d_2=d=odd$ we have
found that the following ``classical'' unitary transformation (which
only permutes the $d^2$ bases states) is  optimal and reaches the bound:
$U\, |i\rangle\otimes |j\rangle = |i+j\rangle \otimes |i-j\rangle
$
where  $ i,j=0 \dots d-1$ and the sums    are mod$\,d.$
[Notice that the above expression for even $d$ does not
define a permutation of the  basis of $\cal H$.]
Similarly a more complicated construction gives an optimal  permutation that
achieves  the bound for the case $d_1=d_2=d=4 n$. Thus for equal
dimensions the only case that remains to be  solved is $d_1=d_2=d=4 n +2$,
e.g. $6\times 6.$ We have also found unitary transformations satisfying the
bound for a very asymmetric case, namely $d_2=n\cdot m$ with $n,m
\ge d_1$. This last example is of the type of controlled unitary operation 
from the larger to the smaller system.
The previous constructions for equal dimensions can be viewed as a
concatenation of two such controlled operations, in which the control role
is played alternatively by one of the subsystems.
It may also be    
 interesting  that for the  cases 2 $\times$ odd the bound
can be shown {\em not} to be reached by permutations.

Let us finally discuss briefly the  numerical evidences. First, even for dimensions  $2\times 3$
the bound appears not to be reached, rather for optimal $U$'s we get the value
 $1/3$ (instead of   $3/8$). For all
other cases that we have checked the bound seems to be reached, namely
for $2\times 4$ up to  $2\times 7$, and $3\times 4$ up to  $3\times 6$.
In conclusion one might conjecture that the only cases where the optimal transformations do not
reach the bound (\ref{bound})  are $2\times 2$ and $2\times 3$ \cite{sure}.

{\em Conclusions.} In this paper we introduced a  measure for the entangling power $e(U)$
of  unitary transformations $U$ acting on the state-space $\cal H$ of a bi-partite $d_1\times d_2$ quantum system.
In terms of this measure we moved a first step towards the analysis of the manifold of bi-partite unitary transformations. 
We  analyzed the induced probability distribution $q(e)$ over ${\cal U}({\cal H})$
as $d_1$ and $d_2$ varies, and we found  an analytical form of optimal transformations
for some cases.
Although we believe that both the questions  addressed and the approach we adopted  are quite natural 
and physically motivated the role, if any,  that the entangling power will play in Quantum Information theory
is still an issue for future work.
 
 The authors thank M. Rasetti and J. Pachos for useful discussions.
Ch. Z. is supported by the EU project IST-Q-ACTA.

\widetext

\begin{references}
\bibitem{sure}
Since our constructions for achieving the bound have a rather
number theoretic character  it is still imaginable that there are more
exceptions at specific higher dimensions.
\end{references}
\end{document}